\documentclass[12pt]{article}

\usepackage{psfig}

\newlength{\dinwidth}               
\newlength{\dinmargin} 
\renewcommand{\arraystretch}{1.2}
\begin{document}


\newcommand{\qq}         {\mbox{$Q^2$}}
\newcommand{\xbj}        {\mbox{$x$}}
\newcommand{\xjet}       {\mbox{$x_{Jet}$}}
\newcommand{\etajet}     {\mbox{$\eta_{Jet}$}}
\newcommand{\eel}        {\mbox{$E_{el}$}}
\newcommand{\ptq}        {\mbox{$p_t^2/Q^2$}}
\newcommand{\xjetxbj}    {\mbox{$ln(x_{Jet}/x)$}}
\newcommand{\GeV}        {\mbox{\rm GeV}}
\newcommand{\GeVc}       {\mbox{\rm GeV/c}}
\newcommand{\kt}         {\mbox{$k_T$}}
\newcommand{\alphas}     {\mbox{$\alpha_S$}}
\newcommand{\ra}         {\mbox{$\rightarrow$}}
\newcommand{\pms}        {\mbox{$\pm$}}
\newcommand{\gap}        {\vspace*{.1cm}}

\def\3{\ss}                                                                                        
\newcommand{\address}{ }                                                                           
\renewcommand{\author}{ }                                                                          


\vspace*{1.cm}

\hfill { \large DESY-98-050}

\vspace*{2.cm}

\begin{center}

{\bf \LARGE  Forward Jet Production in \vspace*{0.3cm} \\
       Deep Inelastic Scattering at HERA }            

\vspace*{1.cm}  
   {\large    ZEUS Collaboration }  \end{center}

\vspace*{6.cm}
\begin{center} \large \bf Abstract \end{center}
\vspace*{0.5cm}
The inclusive forward jet cross section in deep inelastic $e^+p$ scattering
has been measured in the region of $x$--Bjorken,
~$4.5 \cdot 10^{-4}$~ to ~$4.5 \cdot 10^{-2}$. This measurement is 
motivated by the search for effects of BFKL--like parton shower evolution. 
The cross section at hadron level as a function of \xbj\ is compared to 
cross sections predicted by various Monte Carlo models. An excess of
forward jet production at small \xbj\ is observed, which is not reproduced by
models based on DGLAP parton shower evolution. 
The Colour Dipole model describes the data reasonably well. Predictions
of perturbative QCD calculations at the parton level based on BFKL and DGLAP 
parton evolution are  discussed in the context of this measurement.

\thispagestyle{empty}

\newpage

\pagenumbering{Roman}                                                                              
                                                   %
\begin{center}                                                                                     
{                      \Large  The ZEUS Collaboration              }                               
\end{center}                                                                                       
  J.~Breitweg,                                                                                     
  M.~Derrick,                                                                                      
  D.~Krakauer,                                                                                     
  S.~Magill,                                                                                       
  D.~Mikunas,                                                                                      
  B.~Musgrave,                                                                                     
  J.~Repond,                                                                                       
  R.~Stanek,                                                                                       
  R.L.~Talaga,                                                                                     
  R.~Yoshida,                                                                                      
  H.~Zhang  \\                                                                                     
 {\it Argonne National Laboratory, Argonne, IL, USA}~$^{p}$
% ...
\newpage                                                                                           
% ...
$^{\    2}$ now at Univ. of Salerno and INFN Napoli, Italy \\                                      
% ...
$^{  21}$ now at ATLAS Collaboration, Univ. of Munich \\                                           
% ...
$^{  31}$ partially supported by the Foundation for German-Russian Collaboration                   
DFG-RFBR \\ \hspace*{3.5mm} (grant no. 436 RUS 113/248/3 and no. 436 RUS 113/248/2)\\              
% ...
                                                           %
                                                           %
\newpage   
                                                           %
                                                           %
\begin{tabular}[h]{rp{14cm}}                                                                       
$^{a}$ &  supported by the Natural Sciences and Engineering Research                               
          Council of Canada (NSERC)  \\                                                            
$^{b}$ &  supported by the FCAR of Qu\'ebec, Canada  \\                                            
$^{c}$ &  supported by the German Federal Ministry for Education and                               
          Science, Research and Technology (BMBF), under contract                                  
          numbers 057BN19P, 057FR19P, 057HH19P, 057HH29P \\                                        
$^{d}$ &  supported by the MINERVA Gesellschaft f\"ur Forschung GmbH,                              
          the German Israeli Foundation, the U.S.-Israel Binational                                
          Science Foundation, and by the Israel Ministry of Science \\                             
$^{e}$ &  supported by the German-Israeli Foundation, the Israel Science                           
          Foundation, the U.S.-Israel Binational Science Foundation, and by                        
          the Israel Ministry of Science \\                                                        
$^{f}$ &  supported by the Italian National Institute for Nuclear Physics                          
          (INFN) \\                                                                                
$^{g}$ &  supported by the Japanese Ministry of Education, Science and                             
          Culture (the Monbusho) and its grants for Scientific Research \\                         
$^{h}$ &  supported by the Korean Ministry of Education and Korea Science                          
          and Engineering Foundation  \\                                                           
$^{i}$ &  supported by the Netherlands Foundation for Research on                                  
          Matter (FOM) \\                                                                          
$^{j}$ &  supported by the Polish State Committee for Scientific                                   
          Research, grant No.~115/E-343/SPUB/P03/002/97, 2P03B10512,                               
          2P03B10612, 2P03B14212, 2P03B10412 \\                                                    
$^{k}$ &  supported by the Polish State Committee for Scientific                                   
          Research (grant No. 2P03B08308) and Foundation for                                       
          Polish-German Collaboration  \\                                                          
$^{l}$ &  partially supported by the German Federal Ministry for                                   
          Education and Science, Research and Technology (BMBF)  \\                                
$^{m}$ &  supported by the Fund for Fundamental Research of Russian Ministry                       
          for Science and Edu\-cation and by the German Federal Ministry for                       
          Education and Science, Research and Technology (BMBF) \\                                 
$^{n}$ &  supported by the Spanish Ministry of Education                                           
          and Science through funds provided by CICYT \\                                           
$^{o}$ &  supported by the Particle Physics and                                                    
          Astronomy Research Council \\                                                            
$^{p}$ &  supported by the US Department of Energy \\                                              
$^{q}$ &  supported by the US National Science Foundation \\                                       
\end{tabular}                                                                                      
                                                           %
                                                           %
\newpage

\setcounter{page}{1}
\pagenumbering{arabic}                
                                                              

\section{Introduction}
\label{s:intro}
One of the significant discoveries made at HERA was the steep rise of the
proton structure function $F_2(x,Q^2)$ in the region of small
$x$--Bjorken ($\xbj \le 10^{-3}$) \cite{ZEUS_F2, H1_F2}.
Various attempts
have been made to predict the behaviour of $F_2$. In one approach the
$x$ dependence of $F_2$ is fitted at a fixed value of the scale
$Q^2$ and then evolved taking into account $\ln Q^2$ terms 
according to the DGLAP \cite{DGLAP} evolution equations \cite{CTEQ4D,MRS}.
Different starting scales have been chosen for the evolution, down to
values as small as $Q^2\approx 0.3\ {\rm GeV^2}$  \cite{GRV}. In
another approach the leading terms in $\ln (1/x)$ which appear
together with $\ln Q^2$ terms in the evolution equations are resummed
to yield the BFKL equation \cite{BFKL}. The appeal of the BFKL approach 
is the fact that it directly
predicts the behaviour of $F_2$ as a function of $x$. 
The CCFM equation \cite{CCFM} interpolates between the
two approaches and reproduces both the DGLAP and the BFKL behaviour 
in their respective ranges of validity.  From the existing $F_2$
data it is not possible to determine unambiguously whether
the BFKL mechanism plays a significant role in the HERA $x$-range.

\begin{figure}[htb]
\centerline{\psfig{file=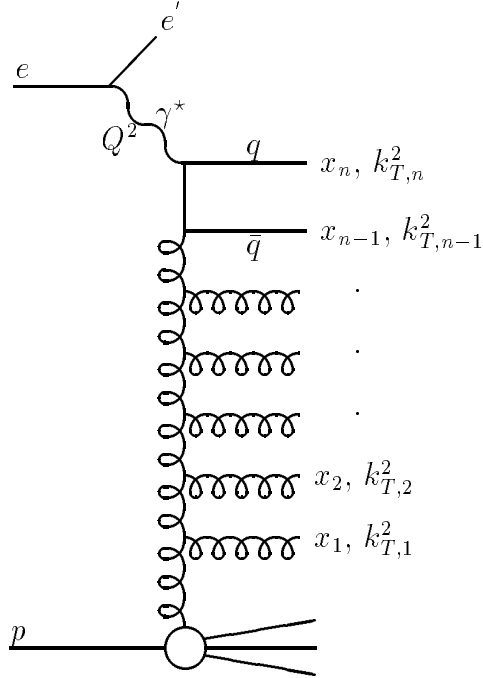,height=9.cm}}
\caption{\it Parton ladder diagram contributing to jet production in DIS. 
The emitted partons in the ladder 
extend from the bottom of the quark box down to the proton. The transverse
momenta, $k_{T,i}$, and the longitudinal momentum fractions, $x_i$, with 
respect to the momenta of the incoming proton, evolve along the ladder.}
\label{fig:gluon_ladder}
\end{figure}

Several exclusive measurements have been proposed in order to 
find evidence for BFKL effects 
\cite{H1_FJETS,H1_Etflow,BFKL_AngDecorr,H1_ChaPar}. 
The method
proposed by Mueller \cite{MUELLER,Bartels} starts from the observation that
the BFKL equation predicts a different ordering of the momenta in the
parton cascade (see figure~\ref{fig:gluon_ladder}).  
The DGLAP equation predicts strong
ordering of the parton transverse momenta $k_{T,i}$ while the BFKL
equation relaxes this ordering but imposes strong ordering of the
longitudinal momenta $x_i$:

\begin{center}
  DGLAP:  \ \ \ $ x = x_n <x_{n-1} < ... < x_1$, \ \ \ \ 
          $ Q^2 = k^2_{T,n} \gg k^2_{T,n-1} \gg ... \gg k^2_{T,1}; $ 

  BFKL: \hspace*{0.7cm} $ x = x_n \ll x_{n-1} \ll ... \ll x_1$, 
       \hspace*{0.5cm}  No \ ordering \ in \ $k_T$. \hspace*{1.5cm}
\end{center}

Thus the BFKL equation predicts additional contributions to the hadronic final
state from partons with large transverse and longitudinal momenta, i.e. high 
transverse momentum partons going forward in the
HERA frame\footnote{
In this paper we use the standard ZEUS right-handed coordinate system, in
which $X=Y=Z=0$ is the nominal interaction point. The positive $Z$-axis
points in the direction of the proton beam, which is referred to as the 
forward direction. The $X$-axis is horizontal, pointing towards the centre
of HERA.}.
These partons may be resolved experimentally as jets and result
in an enhancement of the forward jet cross section at small \xbj\ 
over NLO calculations and parton shower models based on DGLAP evolution.

The H1 collaboration has published results on  the forward jet 
cross section \cite{H1_FJETS} and the
transverse energy flow in the forward direction \cite{H1_Etflow} 
and found trends compatible with the expectations of BFKL dynamics. 

In this paper we study forward jet production in a region which is expected
to be sensitive to BFKL parton dynamics ($E^2_{T,Jet} \approx Q^2$). The
comparison with standard QCD-inspired Monte Carlo models is also
extended to the range where either
$E^2_{T,Jet} \ll Q^2$ or $E^2_{T,Jet} \gg Q^2$.
The measurement of the forward jet cross section is
based on an order-of-magnitude increased statistics compared to 
\cite{H1_FJETS}. 
Comparisons with different Monte Carlo models are presented and theoretical
predictions are discussed in the context of this measurement.
%

\section{Experimental Setup and Trigger}
\label{setup}
In 1995 HERA operated with 174 colliding bunches of $E_p=$ 820~GeV 
protons and $E_e=$ 27.5~GeV positrons. Additionally 21 unpaired 
bunches of protons or positrons circulated to determine the 
beam--related background. 
The integrated luminosity taken in 1995 and used in this analysis is
6.36 pb$^{-1}$.  
A detailed description of the ZEUS detector can be found in \cite{ZEUS}.
In the following the detector components relevant for this analysis are
briefly described.

The central tracking detector (CTD) is surrounded by a superconducting 
solenoid. Outside the solenoid is the uranium calorimeter
\cite{CAL}, which is divided in three parts: forward (FCAL), barrel 
(BCAL) and rear (RCAL) covering the polar angle region of 2.6$^{\rm o}$ to
176.2$^{\rm o}$. The calorimeter covers 99.9\%
of the solid angle, with holes of 20$\times$12 cm$^2$ in the 
centre of the rear and of $20 \times 20$~cm$^2$ in the forward 
calorimeter to accommodate the beam pipe.
Each of the calorimeter parts is subdivided into towers which are segmented
longitudinally into electromagnetic (EMC) and hadronic (HAC) sections.
These sections are further divided into cells, each of which is read out 
by two photomultipliers. The size of the central FCAL EMC (HAC) cells is 
$20 \times 5$~cm$^2$ ($20 \times 20$~cm$^2$). 
From test beam data, energy resolutions of 
$\sigma_E / E = 0.18 / \sqrt{E}$ for positrons and $\sigma_E / E = 0.35 / 
\sqrt{E}$ for hadrons ($E$ in GeV) were obtained.
The effects of uranium noise are minimised by discarding cells in the 
EMC or in the HAC sections if they have energy deposits of less than 60~MeV
or 110~MeV, respectively. 

A three--level trigger is used to select events online~\cite{ZEUS}. 
The selection criteria are similar to those for the measurement of $F_2$ 
\cite{ZEUSF294}. For \qq\ greater than  
10~GeV$^2$ the trigger efficiency for DIS events is above 97\%.
About $7\cdot 10^6$ events survive our trigger requirements.

\section{Event Selection}
\label{ev_sel}
Events are selected off\/line if the $Z$-component of their primary vertex 
reconstructed from tracks in the CTD lies within $\pm 50$~cm around the 
nominal vertex. A positron candidate with an energy $E_{e'}$ above 10~GeV
is required. Candidates hitting the RCAL front face within a box 
of $ X \times Y = 13 \times 8~{\rm cm}^2$ around the beam line are rejected 
because of possible energy loss into the beam hole.
The fractional energy transfer $y_{JB}$ by the virtual photon in the
proton rest frame, calculated from the hadronic energy \cite{jacblo} 
in the calorimeter, is required to be $y_{JB} > 0.1$. This assures sufficient
hadronic energy in the calorimeter to measure the event parameters with
good accuracy. The $y_{el}$, i.e.\ the $y$ parameter calculated from
the positron energy and angle, is required to be below 0.8. 
This rejects photoproduction events with low energy 
fake positron candidates in the FCAL. The quantity 
$E-P_Z = \Sigma_i E_i (1-\cos \theta_i )$ is required to lie 
between 35 and 65~GeV. 
This requirement also removes photoproduction background.
Here $E_i$ and $\theta_i$ are the energies and the polar angles 
of the calorimeter cells.
For DIS events in an ideal detector $E-P_Z$ is 
equal to twice the energy of the incoming positron.
Events within the \xbj\ range of ~$4.5 \cdot 10^{-4}$~ to 
~$4.5 \cdot 10^{-2}$~ are finally selected 
for the cross section measurement. 

\section{Monte Carlo Simulation}
\label{mc}
Monte Carlo event simulation was used to correct the measured 
distributions for detector acceptance and smearing effects. 
The detector simulation is based on the GEANT 
\cite{geant} program and incorporates our understanding of the detector
and of the trigger. 
Events were generated with DJANGO~6.24 \cite{django}, 
which interfaces HERACLES~4.5.2 \cite{HERACLES}
to either the Colour Dipole model \cite{CDM} as implemented 
in ARIADNE~4.08 \cite{ARIADNE} or to LEPTO~6.5 \cite{LEPTO}.
HERACLES includes photon and $Z^0$ exchanges and first order electroweak
radiative corrections.
All samples were generated with the proton structure function 
CTEQ4D \cite{CTEQ4D} which describes the measured $F_2$ structure 
function \cite{ZEUS_F2}.

The Colour Dipole model treats gluons emitted from quark--antiquark pairs 
as radiation from a colour dipole between two partons.
This results in partons which are not ordered in their 
transverse momenta \kt.  Thus ARIADNE is frequently
referred to as ``BFKL--like'' although it does not make explicit use of the
BFKL equation.
The hadronisation in ARIADNE is based on the LUND string model as
implemented in JETSET~\cite{JETSET}. 

Another event sample was generated with the MEPS option of DJANGO, 
as implemented into LEPTO. 
Here the hard interaction is 
taken from the first order matrix element, but the higher 
orders are simulated by a parton shower  based on 
the DGLAP equation. Thus this sample provides a \kt--ordered  
parton shower. The hadronisation of the partons is done
in the same way as in ARIADNE, i.e. via JETSET. In addition to the
hard processes, this version of LEPTO has non--perturbative
effects implemented, so--called Soft Colour Interaction (SCI), 
which affect the investigated phase space region.

A third sample was generated with HERWIG~5.9~\cite{HERWIG},
which like LEPTO has a parton shower evolution based on DGLAP.
However, there are some differences 
with respect to LEPTO, such as the 
implementation of colour coherence effects or gluon splitting 
in the cluster hadronisation of the fragmentation phase.
\newpage
 
A fourth sample was generated at hadron level only with the Linked Dipole 
Chain model option (LDC, version 1.0) \cite{LDC} of ARIADNE,
which uses its own structure functions. The parametrisation of ``set A'' 
was used, which fits data from H1 and ZEUS.
In this model the parton shower evolution is based on 
a reformulation \cite{CCFM_LDC} of the CCFM approach \cite{CCFM} 
which approximates the BFKL 
prediction at low \xbj\ and the DGLAP prediction in the high-\xbj\ limit.

All Monte Carlo events were generated with the default settings of the
input parameters.  
The first three samples were passed through the full trigger and detector 
simulation and were analysed in the same way as the real data.
Additional Monte Carlo samples were generated but not passed 
through the detector simulation in order to study the parton and 
hadron--level properties of the jets.

The effect of initial and final state QED radiation was studied by generating 
ARIADNE events with and without QED corrections. After applying all event
and jet selection cuts the jet cross sections as a function of \xbj\ with and 
without QED corrections agree within 5\%.
We conclude that QED radiation effects are small and ignore them in the 
following.

\section{Jet Finding Algorithm and Jet Selection}
\vspace*{-.5cm}
\subsection{Jet Algorithm}
\label{jet_alg} \vspace*{-.2cm}
The analysis is performed with the cone jet algorithm according to the
Snowmass convention \cite{CONE}. 
The algorithm is applied to calorimeter cells in the laboratory frame 
where the cells assigned to the scattered positron are excluded.

The algorithm maximises the transverse energy flow $E_T$ through a cone
of radius \linebreak
\mbox{$R=\sqrt{(\triangle \eta)^2 + (\triangle \phi)^2}$}. Here $R=1$
is used and $\triangle \eta$ and  $\triangle \phi$ are the differences of 
pseudorapidities and azimuthal angles with respect to the jet direction.
The axis of the jet is calculated as the transverse energy weighted mean
of the pseudorapidity and azimuth of all calorimeter cells belonging to the 
jet. The transverse energy threshold for the seed cells in the algorithm is 
set to 0.5 GeV.  
Two jets are merged if the overlapping energy exceeds 75\% of the total
energy of the jet with the lower energy. Otherwise two jets are formed 
and the common cells are assigned to the nearest jet. 
The massless option, where $E_{T,Jet} = p_{T,Jet}$, is used.

\subsection{Jet Selection Criteria}
\label{jet_sel} \vspace*{-.2cm}
In order to account for the energy loss of the jets in the inactive
material of the detector an energy correction procedure is applied 
both to the total and transverse energies of the jets.
The correction functions were obtained from the  Monte
Carlo simulations and are parametrised as a function of 
the total and transverse jet energy. 

After the energy correction, several cuts are applied to select forward jets.
The pseudorapidity of the jet is restricted to $\eta_{Jet} < 2.6$ 
(equivalent to $\theta_{Jet} > 8.5^{\circ}$). 
In this region the jets are well reconstructed.
The transverse energy of the jets, measured with respect to the
direction of the incoming proton,  is above 5~GeV.
The scaled longitudinal momentum $x_{Jet}=p_{Z,Jet}/820~{\rm GeV}$
is above  0.036. This selects forward jets. The cut 
$0.5<E^2_{T,Jet}/Q^2<2$, together with the $x_{Jet}$ cut,
selects the phase space region where BFKL effects are expected.
In about 2\% of the events, two forward jets survive our selection cuts. 
In these cases the jet with the largest \xjet\ is selected.
All these cuts restrict the \qq\ of the selected events to
values above $\approx 12$~GeV$^2$. 

For events at large values of \xbj, the jet coming from the scattered 
quark can go sufficiently far forward to survive our cuts. 
In order to reject these events, all found 
jets are boosted into the Breit frame \cite{ZEUS_Breit}. 
The boost is calculated from the
four--momentum of the virtual photon, which is taken as the difference between
the incoming and the outgoing four--momenta of the positron.
Those jets which are in the current region of the Breit frame, i.e.\ which
have negative $Z$--momentum,  are rejected.
This only affects the two highest \xbj\ bins where up to 50\% of the 
jets are rejected.

All the selection criteria relevant for the phase space region which defines 
our cross section measurement are listed in table~\ref{cuts}.
A total of 2918 events with forward jets survive these cuts.

\vspace*{0.3cm}
\begin{table}[htb]
\begin{center}
\begin{tabular}{c}
\hline
\hline
   $E_{e'} >  10$~GeV         \\
   $y > 0.1 $                 \\
   $\eta_{Jet} < 2.6$         \\   
   $E_{T,Jet}> 5$~GeV         \\ 
   $x_{Jet} > 0.036$          \\ 
   $0.5 < E^2_{T,Jet}/Q^2< 2$ \\ 
   $p_{Z,Jet}(Breit) > 0$     \\ 
   $4.5\cdot 10^{-4} < x < 4.5 \cdot 10^{-2}$  \\ 
\hline
\hline
\end{tabular}
\caption{\it Selected phase space region for the cross section measurement.}
\label{cuts}
\end{center}
\end{table}  
\vspace*{-0.5cm}

\section{Comparison of Data and Monte Carlo Distributions}
\label{resol}
Before applying the jet selection criteria, the data and the Monte Carlo 
predictions are compared. They agree well in absolute normalisation and in 
shape.

\begin{figure}[htb]
{\hspace*{0.3cm} \psfig{file=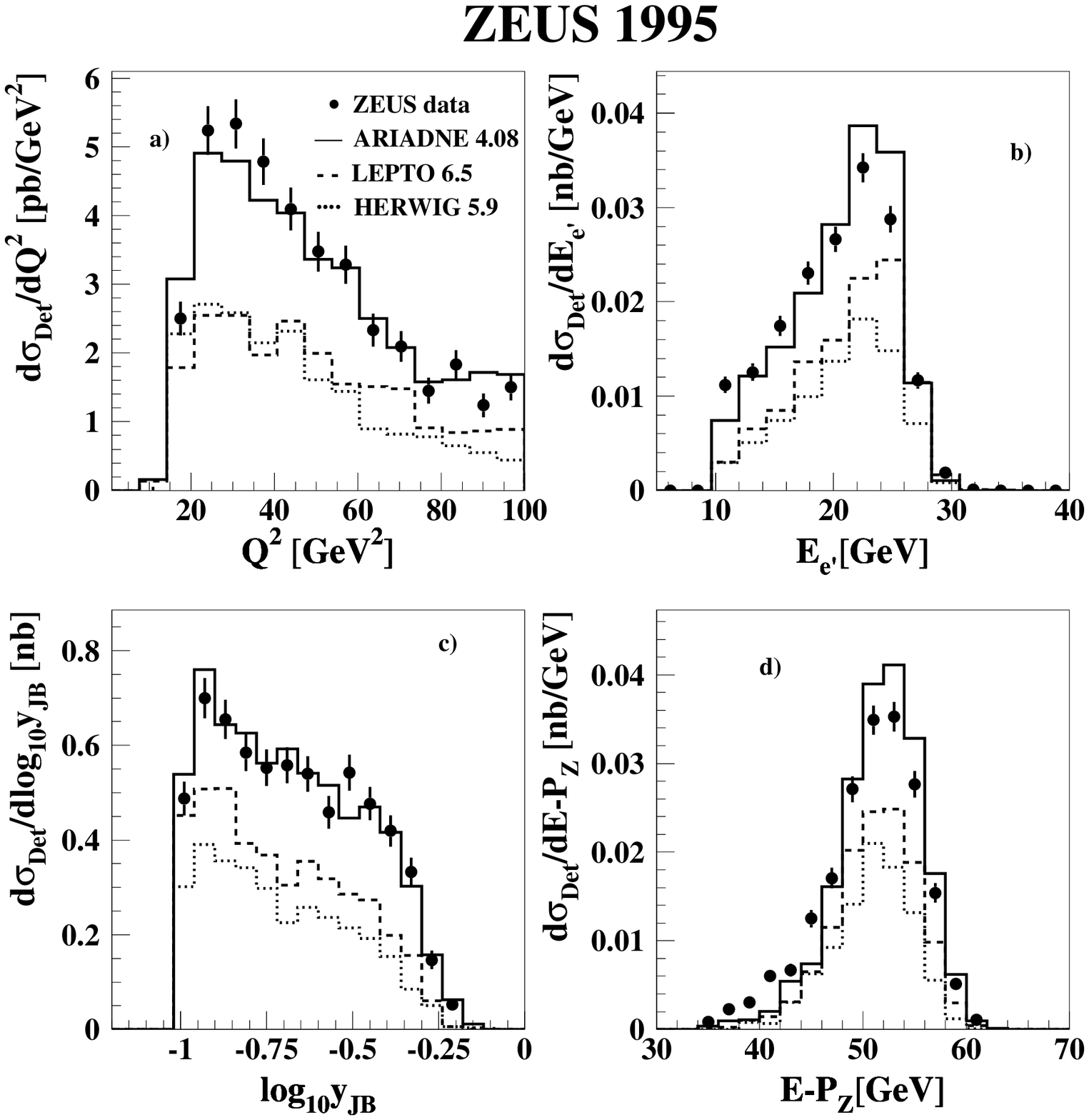,width=18.cm}}
\caption{\it Uncorrected detector--level cross sections after the
             event and jet selection cuts as a function of 
             a) \qq, b) $E_{e'}$, c) $y$, d) $E-P_Z$.
             Data are shown as points, ARIADNE as the full histogram, 
             LEPTO as the dashed histogram and HERWIG as the 
             dotted histogram. Only statistical errors are shown.}
\label{fig:shape_kin}
\end{figure}

After applying the jet selection cuts the uncorrected differential cross 
sections as a function of the event related quantities $Q^2$, $E_{e'}$, 
$y_{JB}$ and $E-P_{Z}$ are compared in figure~\ref{fig:shape_kin}. 
ARIADNE describes the measured distributions reasonably well while the
LEPTO and HERWIG cross sections are too small. 

\begin{figure}[hbt]
{\hspace*{0.3cm} \psfig{file=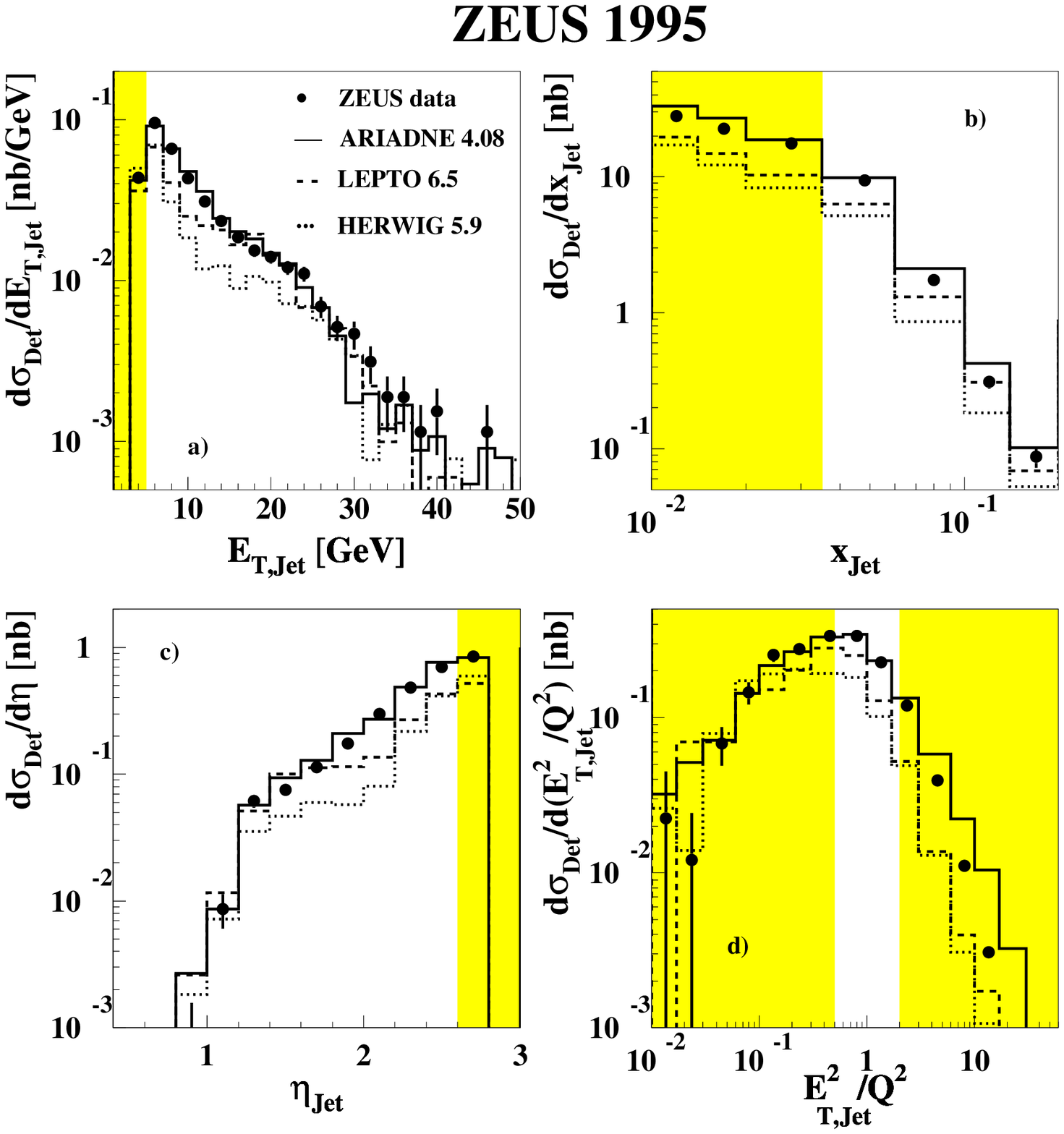,width=18.cm}}
\vspace*{-.5cm}
\caption{\it Uncorrected detector--level cross sections are shown
       as a function of
       a) $E_{T,Jet}$,        
       b) $x_{Jet}$,          
       c) $\eta_{Jet}$,       
       d) $E^2_{T,Jet}/Q^2$.  
       All event and jet selection cuts are
       applied (see text) apart from the cut on the displayed
       variable. Data are shown as points, ARIADNE as the full histogram, 
       LEPTO as the dashed histogram and HERWIG as the dotted histogram.
       The data in the shaded areas are excluded from the final cross 
       section measurement. Only statistical errors are shown.}
\label{fig:effpar_jet}
\end{figure}

In figure~\ref{fig:effpar_jet} we show uncorrected detector--level cross 
sections as a function of the jet--related quantities
$E_{T,Jet}$, $x_{Jet}$, $\eta_{Jet}$ and $E_{T,Jet}^2/Q^2$.  
The distributions are compared to the predictions 
of the various Monte Carlo models. All selection criteria  
are applied except the one for the displayed variable. 
The data in the shaded areas are excluded from the final cross 
section measurement.
ARIADNE describes the data in the first three distributions both
in shape and in absolute value over the entire range.
HERWIG and LEPTO underestimate the cross
section significantly and also disagree in shape.

The distribution in figure~\ref{fig:effpar_jet}d) 
can be subdivided into three regions. For small $E^2_{T,Jet}/Q^2$, i.e.
the classical DIS regime, all three models agree with the data. 
Here \qq\ is large compared to $E^2_{T,Jet}$ and the DGLAP--based Monte 
Carlo models are expected to describe the data.
In the unshaded region, which is selected for this analysis, only ARIADNE 
reproduces the data. HERWIG and LEPTO predict much smaller cross
sections. In this area we expect significant contributions from BFKL--based
parton showers.
For higher values of $E^2_{T,Jet}/Q^2$ no model agrees with the \linebreak
%
%
%
data. The cross section of ARIADNE is too large, whereas those of LEPTO and 
HERWIG are too small. 
In this regime the hard scale is no longer given by the invariant mass 
squared, $Q^2$, of the virtual photon, but by the $E^2_T$ of the jets. 

A study of energy flows similar to \cite{ZEUS_JETshapes} has been performed 
in order to investigate whether jets
in the forward region of the detector still have a pronounced signature 
and how the beam hole and the proton remnant affect the selected jets.

\begin{figure}[t]
\centerline{\psfig{file=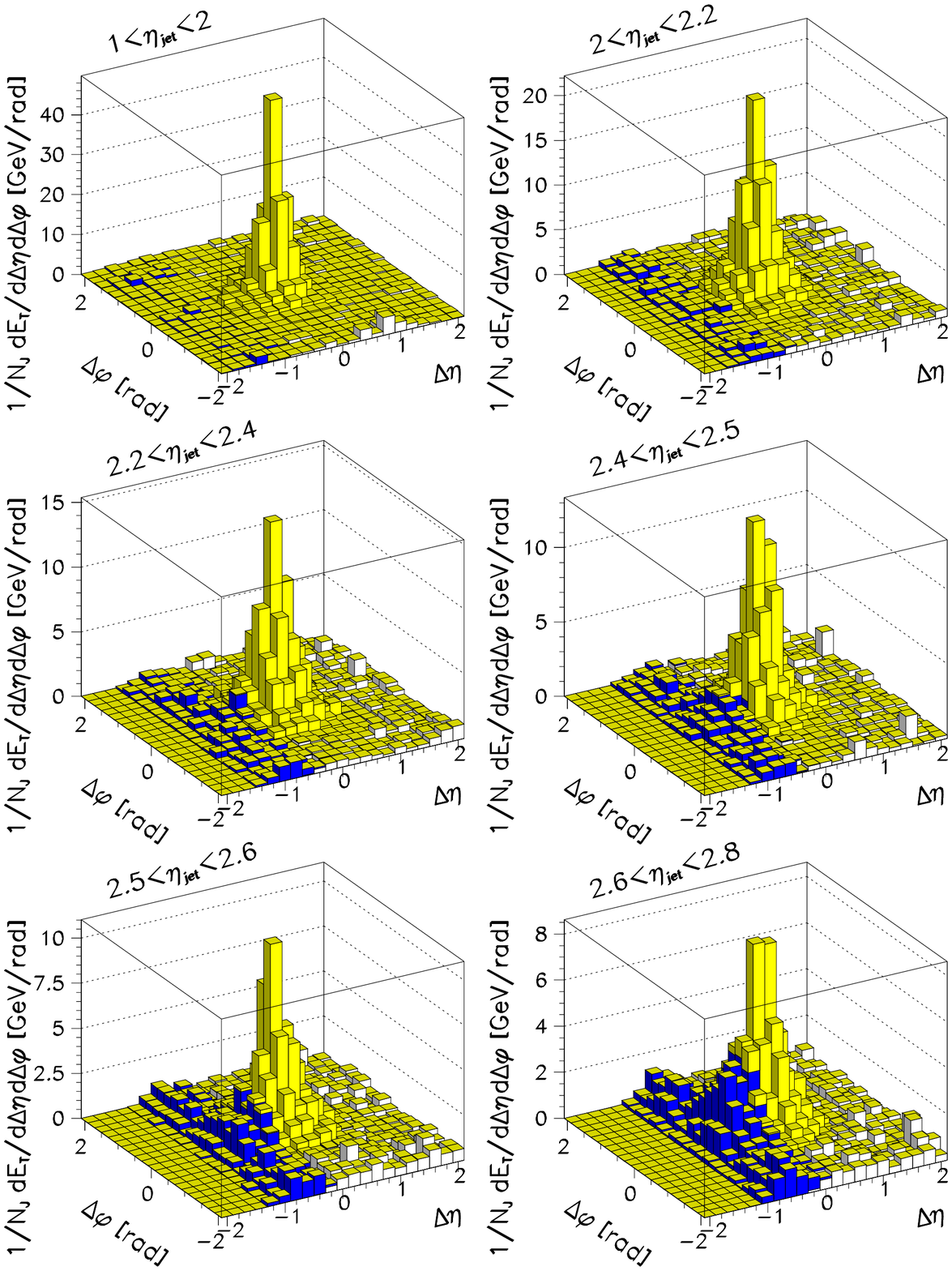,height=22.cm,width=15.cm}}
\vspace{-1.5cm}
\caption{\it The transverse energy flow around the forward jet axis
  averaged over all selected forward jets for various $\eta_{Jet}$ regions. 
  Grey bars indicate the transverse energy in the calorimeter 
  cells attributed to the jet, black bars correspond to the transverse 
  energy in the cells around the beam hole, and white 
  bars correspond to the transverse energy elsewhere in the calorimeter.}
\label{fig:remn7f2}
\end{figure}

In figure~\ref{fig:remn7f2} we show the transverse energy flow with 
respect to the forward jet axis averaged over all selected events 
as a function of 
$\triangle \phi$ and $\triangle \eta$, the difference in azimuth and 
pseudorapidity of the cells with respect to the jet direction.
The grey bars indicate energy deposits in cells which are attributed 
to the forward jet. The black bars indicate the contributions from 
those cells situated in the towers directly surrounding the forward beam 
hole. As can be seen, some black bars also belong to the jet.
The white bars indicate energy deposits which belong neither to the jet
nor to the cells surrounding the FCAL beam hole. For increasing $\eta_{Jet}$  
the black band--like structure in the forward direction of the jet, 
which we attribute to the proton remnant,
becomes more and more prominent. 
For $\eta_{Jet} > 2.6$ the selected jets pick up significant contributions 
from the proton remnant in their tails. Studies of the 
reconstruction accuracy of the angle and of the energy of the jets 
also show a degradation at $\eta_{Jet}$ values above 2.6 (not shown).
Therefore, we require $\eta_{Jet}<2.6$ for this analysis.

\begin{figure}[htb]
\centerline{\psfig{file=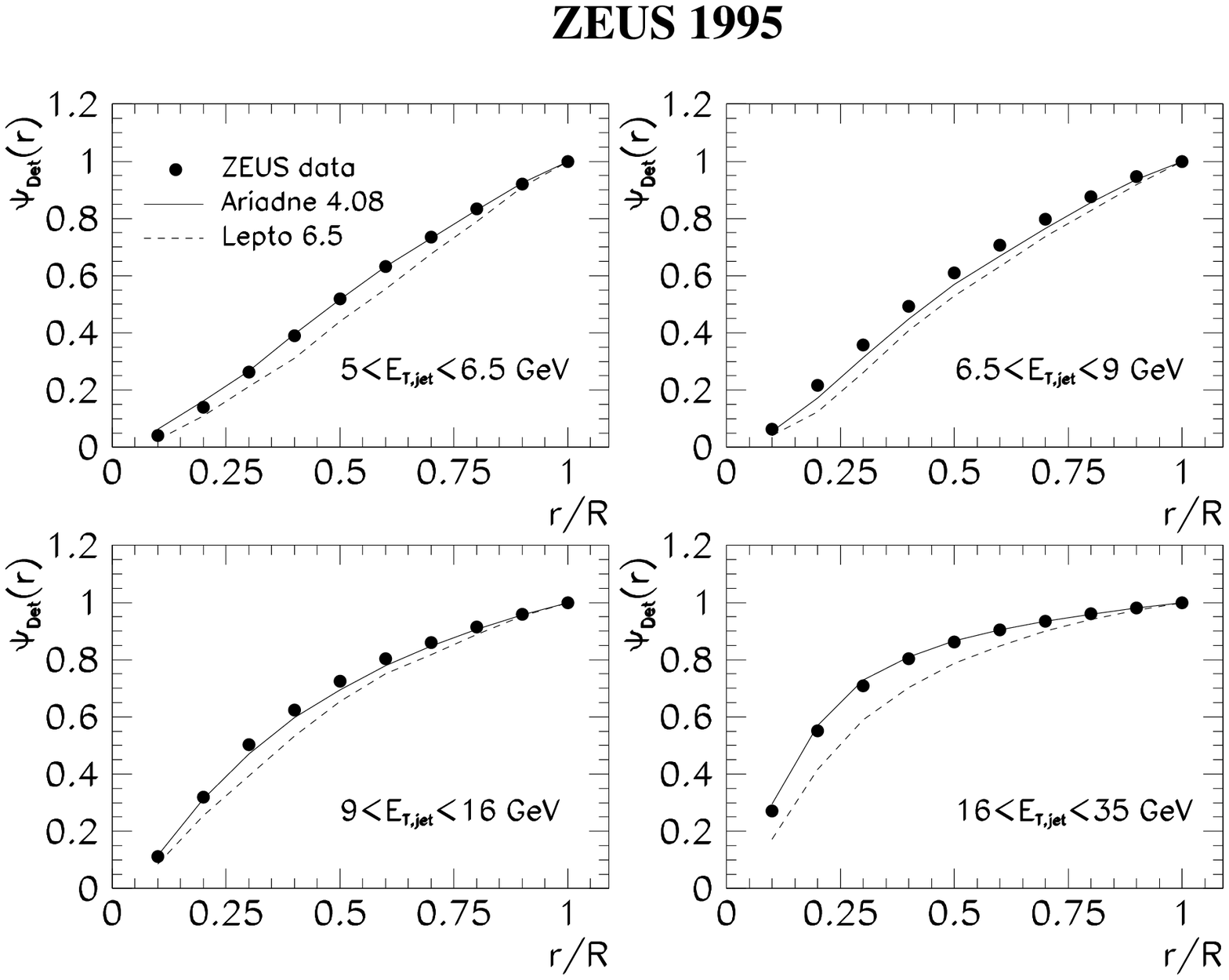,width=18.cm}}
\caption{\it Uncorrected integrated jet shapes in intervals of $E_{T,Jet}$
              which are chosen to contain approximately equal numbers of 
              jets.   Data are shown as dots. 
     Predictions from  ARIADNE (LEPTO) are shown as full (dashed) lines.
     All event and jet selection cuts are applied. The statistical errors 
     are negligible compared to the size of the dots.}
\label{fig:fj_int_shapes}
\end{figure}

In figure~\ref{fig:fj_int_shapes} we show the integrated jet shape
$\Psi_{Det}(r)$, i.e. the relative amount of transverse energy deposited 
inside a cone of radius $r<R$ with respect to the jet axis.
This function is defined as 
$$ \Psi_{Det}(r) = \frac{1}{N_{Jets}} \sum_{Jets} \frac{E_T(r)}
              {E_T(r=R)},$$
where $E_T(r)$ is the sum of the transverse energies of all cells of 
a given jet within a radius $r$ with respect to the jet axis. 
ARIADNE  describes the distribution well for all values of $E_{T,Jet}$.
LEPTO generates broader jets than observed in the data.
The jets are more collimated as $E_{T,Jet}$ increases.
A similar level of agreement between the data and
the Monte Carlo events is found when bins of $\eta_{Jet}$ instead of 
$E_{T,Jet}$ are investigated (not shown).

\section{Jet Finding Efficiencies and Purities}
\label{effi}

A detailed study of the jet reconstruction quality was performed in order 
to find acceptable cuts for the analysis.

ARIADNE was used for the study of the efficiencies and purities of the
jet reconstruction and of the acceptance correction since it
describes the data best in shape and absolute rate.
The efficiency $\epsilon$ and purity $p$ of the jet finding are determined
as a function of \xbj\ and are defined by:

$$ \epsilon = { {Number~of~jets^{det \oplus had}} \over 
                {Number~of~jets^{had}} }, 
 \hspace*{3.cm} 
p = { {Number~of~jets^{det \oplus had}} \over 
      {Number~of~jets^{det}} }. $$

\begin{figure}[htb]
\centerline{\psfig{file=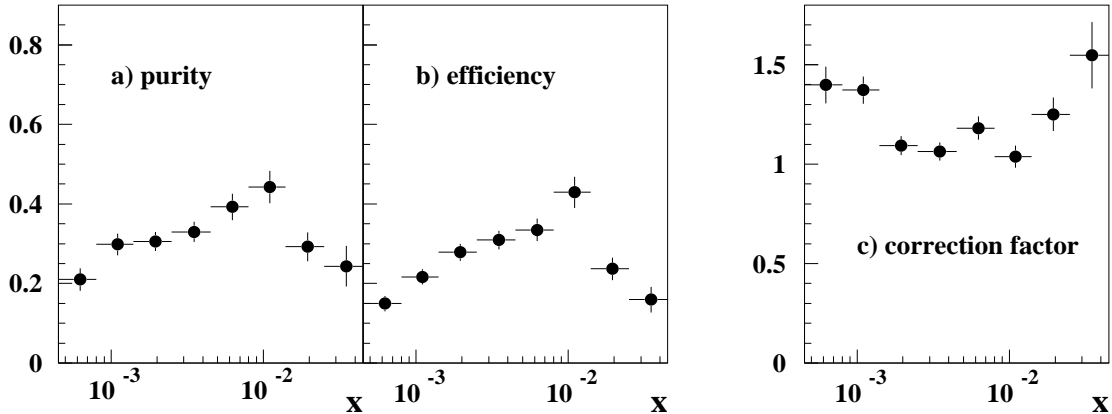,width=17.cm}} 
\caption{\it a) Purity, b) efficiency and c) correction 
factors for the correction of the detector--level jet cross section to the
hadron--level jet cross section as a function of \xbj.}
\label{fig:effipuri}
\end{figure}

The indices {\it det\/} and {\it had\/} correspond to jets found at the 
detector or hadron level, respectively.  
Hadron--level jets are defined as jets found by the jet algorithm when 
applied to the stable hadrons from the event generator.
The symbol \ {\it det\/}$\oplus${\it had\/} \ means that the jet has to be 
found at both \ levels in the \ relevant \ variable \ range and in the same 
\xbj--bin. The {\it det\/} jets \ are \ those  \newpage
\clearpage
%
surviving all the event and jet selection
criteria and the {\it had\/} jets have to survive only those 
cuts which define 
the phase space region for which the final forward jet cross section is given,
see table~\ref{cuts}. 
Figure~\ref{fig:effipuri} shows as a function of \xbj\ the efficiencies 
$\epsilon$, purities $p$ and correction factors $c=p/\epsilon$ 
for the acceptance correction from detector to hadron level.
The \xbj\ bins are chosen such that the bin width is at least 2--3~times
the \xbj\ resolution and the statistical errors are below 20\%.
The values for $\epsilon$ and $p$ lie between 20\% and 50\%, while those 
of the correction factors are between 1.0 and 1.5.
The drop in $\epsilon$ and $p$ for $\xbj > 10^{-2}$ is due to the 
degraded resolution of \xbj\ in this region of \xbj.
The small overall values of $\epsilon$ and $p$ are a result of the jet 
selection cuts and of the finite resolutions of the jet variables. 
The resolutions are: $\Delta E_{T,Jet} / E_{T,Jet} \simeq 10$\%,
$\Delta \eta_{Jet} \simeq 0.1$, $\Delta x_{Jet}/x_{Jet} \simeq 11$\%, and 
$\Delta (E^2_{T,Jet}/Q^2)~/~(E^2_{T,Jet}/Q^2) \simeq 25$\%. The latter
has the largest effect on $\epsilon$ and $p$.
When the cut on $E^2_{T,Jet}/Q^2$
is dropped the efficiencies and purities increase by about a
factor of two. 
The measured detector--level rates are multiplied bin--by--bin using the
correction factor $c$ 
in order to obtain the hadron--level distributions.
In an independent analysis the acceptance correction was evaluated
using the Bayes unfolding method~\cite{BAYES}. This method \newpage
%
%
takes migration effects between the bins properly into account and 
is a cross check of the reliability of the bin--by--bin acceptance 
correction. The extracted forward jet cross sections agree well between 
the two methods. 

The analysis was also repeated using the \kt\ clustering 
algorithm~\cite{ktalg} in the Breit frame
with $Q^2$ as scale and a resolution parameter $y_{cut} = 0.5$.
With these settings the algorithm creates a large number of jets at the
detector level, which do not have a corresponding jet at the hadron or parton 
level. A change of the $y_{cut}$--parameter does not improve this situation.
Purities and efficiencies in the lowest \xbj\ bins 
are therefore very small, around 5\% and 15\%, respectively. 
Nevertheless, the \kt\ analysis leads to conclusions consistent 
with those drawn in this paper.

\section{Comparison of DGLAP and BFKL Approaches}
%
Perturbative QCD predicts the dynamics of the parton evolution. 
The conventional method to solve the parton evolution equations 
is the DGLAP approach \cite{DGLAP}, which resums leading order 
(LO) terms proportional to $(\ln Q^2)^n$. This results in a parton
cascade strongly ordered in transverse momentum \kt,
where the parton with the highest transverse momentum appears at the 
lepton vertex of the chain. The longitudinal momenta \xjet\ decrease 
towards the photon vertex. 

Next--to--leading--order (NLO) calculations,  i.e.\ full second order 
matrix element calculations including first order virtual corrections, 
where parton densities are incorporated according to the DGLAP scheme, 
are available in three program packages MEPJET, DISENT and DISASTER$++$
\cite{MEPJET,DISENT,DISASTER}. 
The programs use different techniques to calculate cross sections 
but nevertheless agree reasonably well in their predictions \cite{DISASTER}.
These NLO calculations are not available in full Monte Carlo
event simulations. They are purely parton--level calculations,
delivering parton four momenta which can be analysed, such that the parton 
level cross sections can be evaluated using different jet algorithms and 
recombination schemes.  
\newpage

The BFKL approach \cite{BFKL} of the parton evolution resums terms 
proportional to $(\ln 1/\xbj)^n$, which become dominant over 
the $\ln Q^2$ terms at small \xbj. This approach is expected to be valid 
in the high energy limit, where the total available energy, $W$, is large 
with respect to any other hard scale, $E_{T,Jet}$ or $Q$, in DIS. 
The first term of this resummation is second order in the
strong coupling constant $\alpha_s$ and is therefore included in the 
next--to--leading order tree--level diagrams in DGLAP--based calculations,
e.g.\ in MEPJET. In figure~\ref{fig:gluon_ladder} this term corresponds to 
exactly one parton rung
in the gluon ladder between the quark box and the proton.
In the following it will be referred to as the BFKL 1$^{st}$ term.
Since the present approach is only leading $\ln 1/x$,
the parton emissions are strongly ordered in $x_{Jet}$. 
Recent calculations of next--to--leading $\ln 1/\xbj$ terms in 
the BFKL kernel \cite{BFKL_NLO} predict large negative corrections due to
a weakening of the strong ordering in $x_{Jet}$  which
are expected to reduce the cross section significantly. 

The present BFKL calculations do not allow the implementation of a jet 
algorithm. Therefore this calculation can only be regarded as approximate, 
since the measured jet rates depend on the jet algorithm and on their 
resolution parameters, scales and recombination schemes.
For example, in our selected phase space region MEPJET yields cross sections 
for the cone and the \kt\ algorithm with their particular choice of 
parameters which differ by up to 15\%.

\begin{figure}[hbt] 
    \vspace*{-0.5cm}
    \centerline{\psfig{figure=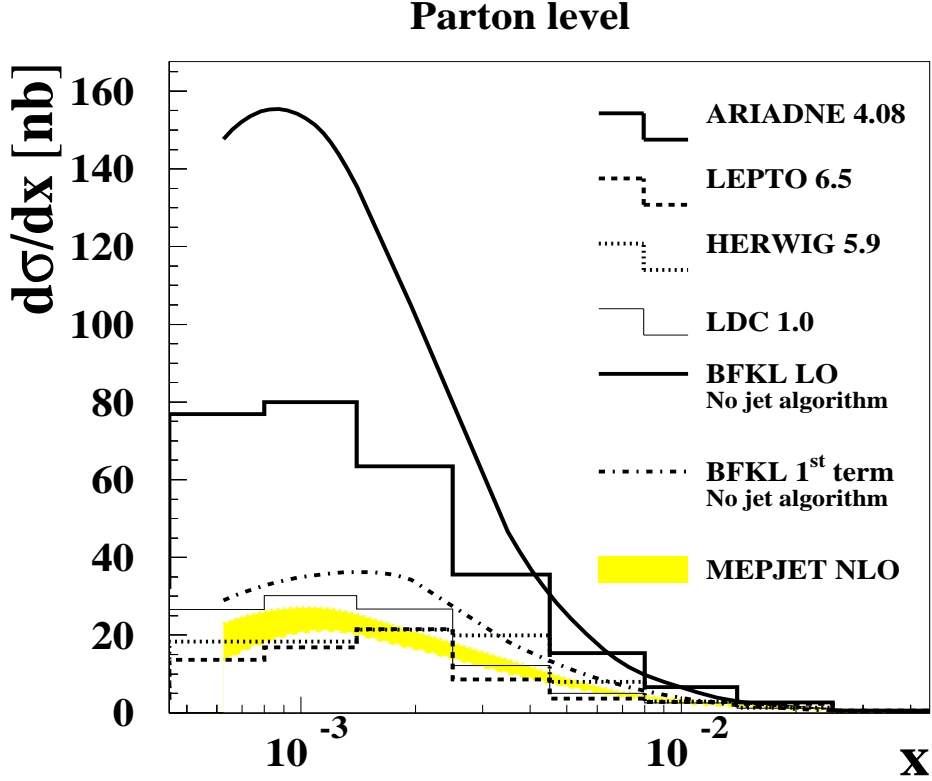,width=14.cm,height=13.0cm}} 
    \vspace*{-2.1cm}
    \caption{\it 
    Parton--level predictions for forward jet cross sections as a function of 
    \xbj. The LO BFKL calculation is indicated by the full curve and the
    BFKL 1$^{st}$--term calculation as the dashed--dotted curve.
    The shaded band gives the range of results obtained with MEPJET using 
    a renormalisation scale between $0.25~k^2_T$ and $2.0~k^2_T$.
    The predictions from ARIADNE (full histogram), LEPTO (dashed histogram), 
    HERWIG (dotted histogram) and LDC (thin full histogram) are also 
    shown.}
    \label{fig:parthad}
\end{figure}

In figure~\ref{fig:parthad} we compare the differential forward jet cross 
section prediction from MEPJET to  
the BFKL 1$^{st}$ term and to the leading order (LO) BFKL calculation.
We have applied the cone algorithm within MEPJET, the same algorithm as used 
for the data.
The renormalisation scale in the MEPJET program is 
varied between 0.25~$k^2_T$ and  2~$k^2_T$
to study the scale dependence of the parton--level cross section. 
Here $k_T$ is the scalar sum of the transverse momenta of the jets in the
Breit frame.
The result changes by $\sim$30\% in
the two small--\xbj\ bins and less than 10\% in the other bins.
This is indicated by the shaded band in figure~\ref{fig:parthad}.

The MEPJET NLO and the BFKL 1$^{st}$--term calculations are similar and
predict a much smaller cross section than the LO BFKL calculation, which
shows a steep rise towards smaller values of \xbj.

Also shown are the parton--level
cross sections from LEPTO, HERWIG, ARIADNE and from LDC.
Both the MEPS--based LEPTO and HERWIG models show reasonable agreement with 
the MEPJET calculations, whereas ARIADNE exhibits a stronger 
increase of the cross section for small \xbj.
The LDC model is well below the ARIADNE predictions.
For increasing \xbj\ all models and calculations converge.

\begin{figure}[hbt]
     \vspace*{-0.5cm}
     \centerline{\psfig{figure=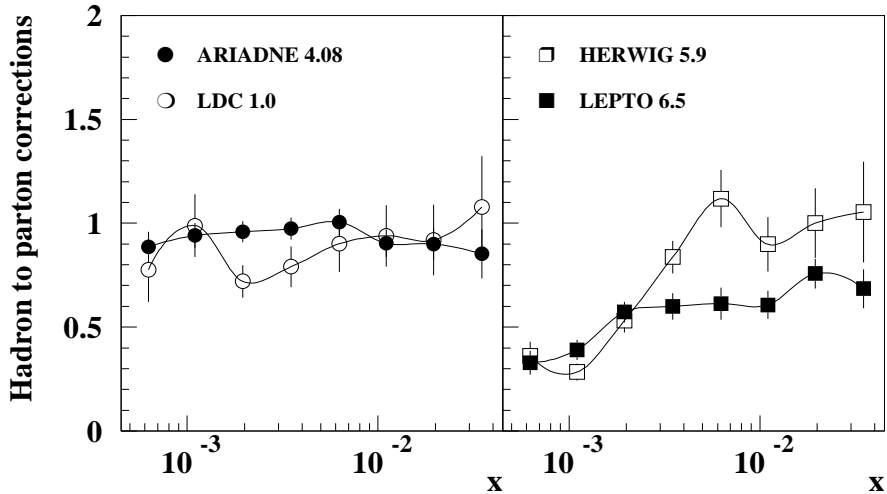,width=14.cm}} 
     \vspace*{-2.5cm}
     \caption{\it Correction factors corresponding to the ratio of the parton
                  level to the hadron--level forward jet cross section as a
                  function of \xbj.
                  The factors for ARIADNE are shown as full circles,
                  for LDC as open circles, for HERWIG as open rectangles 
                  and for LEPTO as full rectangles. }
    \label{parthad_corr}
\end{figure}

For a direct comparison of the data to the theoretical calculations,
the measurements need to be evaluated at the parton level, where
partons are defined in the Monte Carlo at the stage after the last 
branching of the parton shower and before the hadronisation.
The size of the corrections from hadron to parton level is studied 
using the Monte Carlo simulation programs and are
displayed in figure~\ref{parthad_corr}. ARIADNE yields factors
close to unity and shows no dependence as a function of $x$.
LEPTO and HERWIG show large corrections for small $x$ values. 
This is expected, because these DGLAP based models, which have LO matrix 
element calculations implemented, can only produce a significant number of 
forward jets due to hadronisation effects and detector smearing.
The LDC corrections are intermediate to LEPTO/HERWIG and ARIADNE.
As shown above LEPTO and HERWIG also fail to 
describe the cross section both in absolute size and in shape.
Furthermore, the relation between the parton level in parton shower 
Monte Carlo programs and partons in exact NLO calculations is not obvious.
Therefore we refrain from quoting measurements corrected to the parton level. 
%
\clearpage

\section{Systematic Studies}
\begin{figure}[htb]
\centerline{\psfig{file=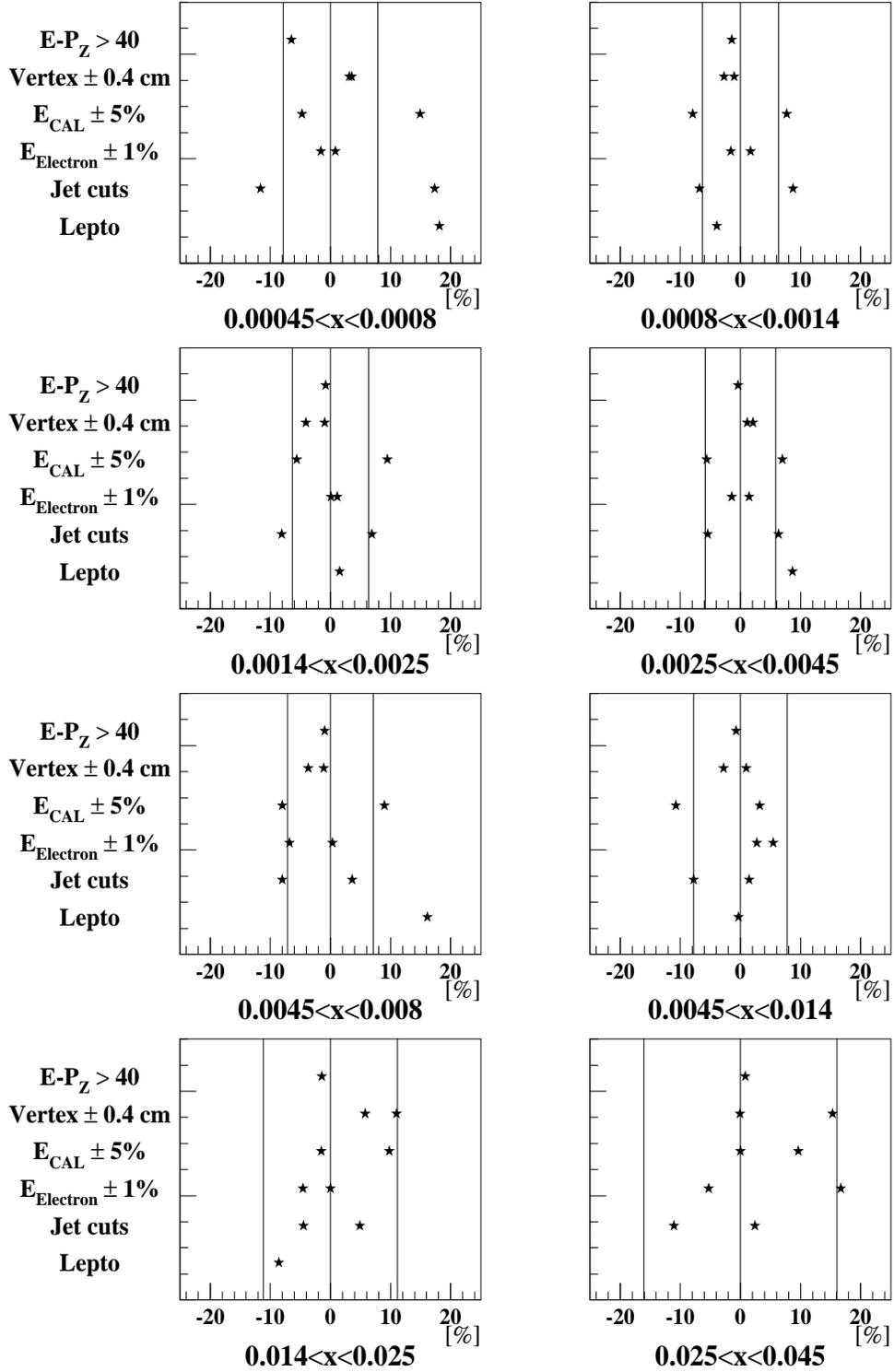,height=22.cm,width=15.cm}}
\caption{\it Percentage change of the forward jet cross section
             due to various systematic checks in the different \xbj\ bins.
             The vertical lines indicate the 
             statistical errors in each interval. Empty bins contain values 
             outside the displayed range (see text).}
\label{fig:systematics}
\end{figure}

\vspace*{-0.3cm}
\label{sys}
We have studied the effects of the variation of several selection cuts and
reconstruction uncertainties on the final cross section. 
Figure~\ref{fig:systematics}
shows the percentage change of the final cross section in each 
\xbj\ interval for the major contributions to the overall systematic error.

{\bf Change of the $E-P_Z$ cut from $>$~35 to $>$~40~GeV} \newline
     This tests the amount of photoproduction background in the 
     sample and changes the result by less than 6\%.

{\bf Alignment uncertainty between the CTD and the FCAL} \newline
      The primary event vertex is determined from tracks in the CTD.
      In order to account for an alignment uncertainty between CTD and
      FCAL, the $Z$--position of the vertex is shifted by $\pm$ 0.4~cm.
      This affects mostly parameters calculated for the scattered
      positron and related quantities like \xbj, \qq\ and the
      boost to the Breit frame. The uncertainty from this effect
      is around 5\%, except in the highest \xbj\ bin where
      it reaches 14\%.  Here, due to the misreconstruction of the 
      kinematic variables the current jet may be reconstructed 
      sufficiently far forward to survive our selection cuts.

{\bf Jet energy scale uncertainty} 
     \newline 
      The energy of the jets is scaled by $\pm$5\% in the Monte Carlo,
      reflecting a global uncertainty of the hadronic energy
      scale in the forward region of the FCAL. The result changes by 
      less than 15\%.

{\bf Electromagnetic energy scale uncertainty of the calorimeter}
      \newline
      The energy of the scattered positrons in the RCAL 
      is scaled by $\pm$1\% in the
      Monte Carlo corresponding to the global uncertainty of the
      electromagnetic energy scale in the calorimeter.
      The result changes typically by less than 5\%.
 
{\bf Uncertainty from jet selection criteria } \newline
  A possible mismatch between the distributions of jet variables in the data
  and in the Monte Carlo is tested by a variation of the cut values by
  about one sigma of their resolution followed by the determination of the 
  cross sections at the nominal value of the cuts at the hadron level. 
  The tested cuts are:
  the minimum $E_{T,Jet}$ (changed from 5~GeV to 4.5 and 5.5~GeV),
  the minimum \xjet\ (changed from 0.036 to 0.042 and to 0.030), 
  the maximum $\eta_{Jet}$ (changed from 2.6 to 2.7 and 2.5), and the 
  minimum and maximum  $E^2_T/Q^2$ (changed from 0.5 to 0.6 and 0.4 or
  from 2 to 2.4 and 1.6, respectively). All these effects are at the 
  level of a few percent, except in
  the lowest \xbj\ bin, where they add up to 18\%. They are
  combined (i.e.\ added in quadrature separately for the positive and negative
  changes) and shown as  ``jet cuts'' in figure~\ref{fig:systematics}.

{\bf Acceptance correction with LEPTO} 
     \newline 
  The full acceptance correction is performed with LEPTO instead of ARIADNE. 
  Since the $E^2_{T,Jet}/Q^2$ distribution of LEPTO differs substantially
  from the measured one, the LEPTO events 
  were reweighted to reproduce the observed $E^2_{T,Jet}/Q^2$ distribution. 
  This test changes the cross section by less than 15\%, except in the
  lowest and the highest $x$ bins, where it increases the result
  by +20\% and +60\%, respectively.

Further checks included the variation of the accepted ranges of the
primary event vertex
($\pm 40$~cm or $\pm 60$~cm instead of $\pm 50$~cm), 
a change of the region for the rejected scattered positrons within a box of
$13 \times 8$~cm$^2$ or $14 \times 14$~cm$^2$ around the RCAL beam pipe,
and a variation of the $y_{el}$ cut from $<$~0.8 to $<$~0.95.
The effect on the cross section is negligible.
A global uncertainty of 1.5\% coming from the luminosity measurement 
is not included.
%
\clearpage

\section{Hadron--Level Forward Jet Cross Section}
\label{results}
\begin{figure}[htb]
\centerline{\psfig{file=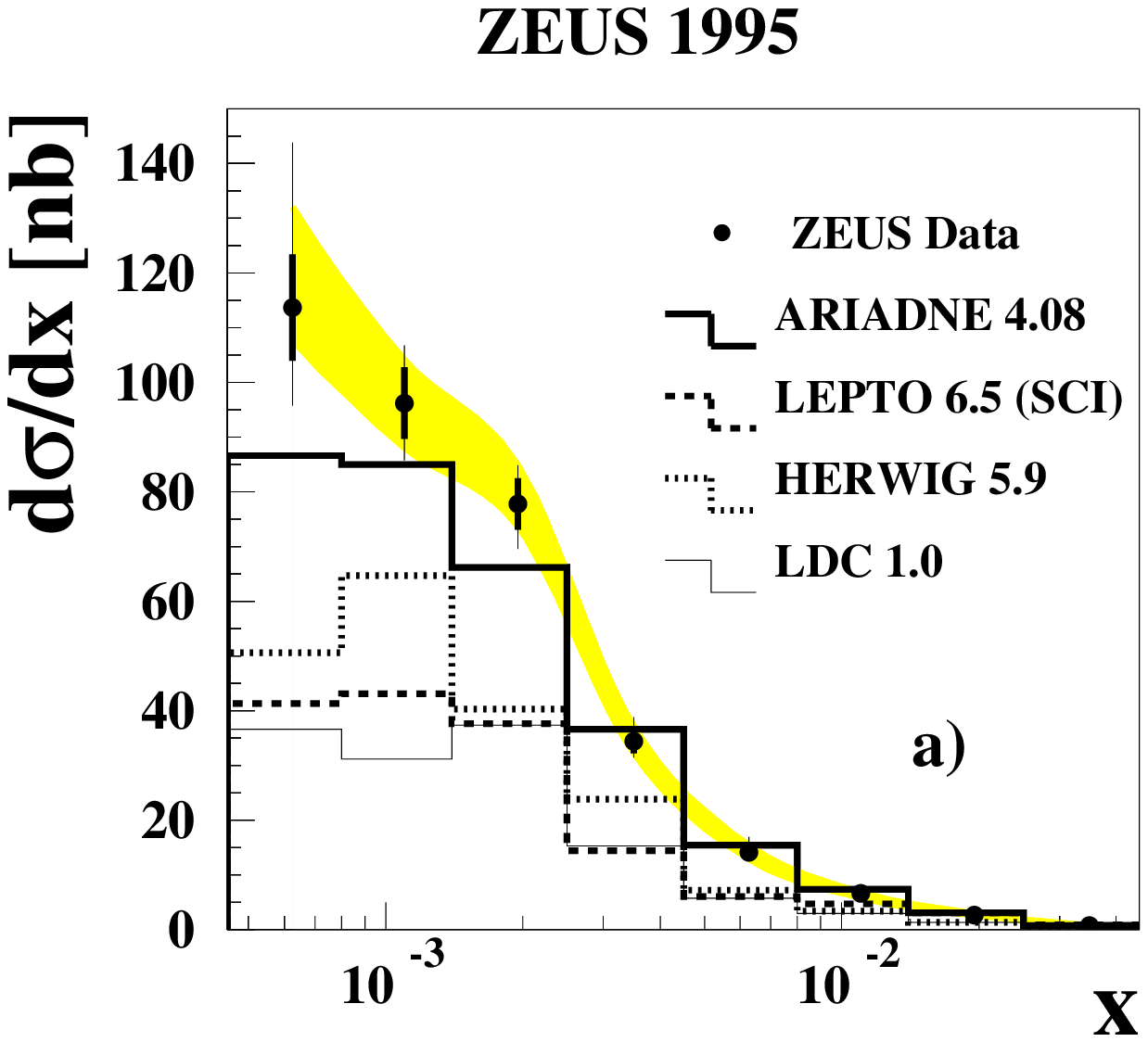,width=13.0cm}}
\vspace{-2.5cm}
\centerline{\psfig{file=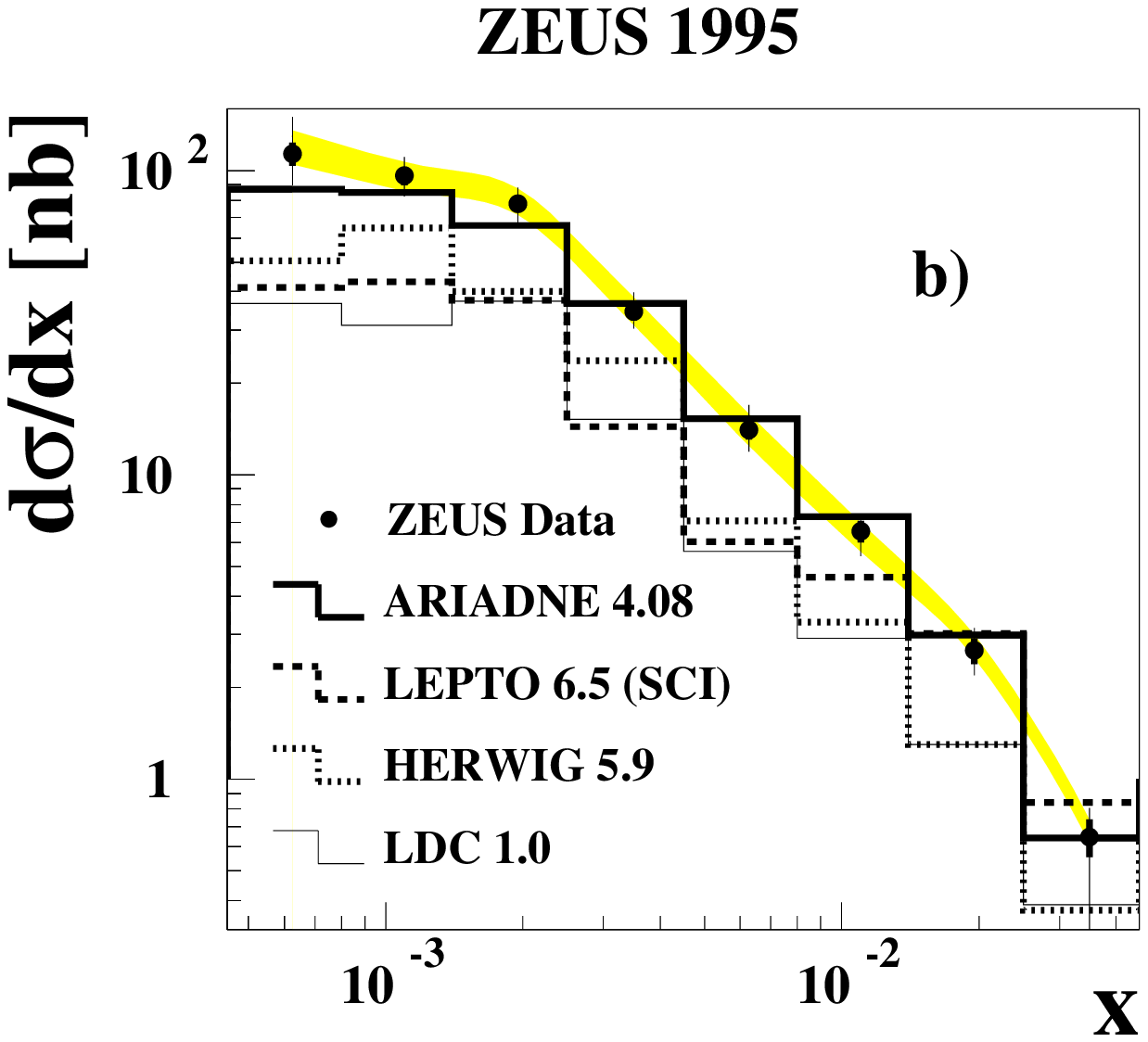,width=13.0cm}}
 \vspace{-2.6cm} 
\caption{\it Forward jet cross section at  hadron level as a
         function of \xbj\ in the kinematic region 
         \etajet$<2.6$, \xjet$>0.036$, $0.5<E^2_{T,Jet}/Q^2<2$,
         $E_{T,Jet}>5$~GeV, $E_{e'} > 10$~GeV, $y>0.1$.
         a) Linear scale, b) logarithmic scale.
         Statistical errors are shown as thick error bars, and statistical and 
         systematic errors added in quadrature as thin error bars. 
         The errors due to the uncertainty of the jet energy scale
         are shown as the shaded band.}
\label{had_ari}
\end{figure}

In figure~\ref{had_ari} we present the hadron--level forward jet cross 
section. The numerical values are given in table~\ref{tab_results}.
ARIADNE describes the hadron--level forward jet cross section reasonably
well, apart from the small--\xbj\ region, where it is slightly below the data. 
LEPTO and HERWIG, as well as LDC, predict significantly smaller cross 
sections. 
Switching off the Soft Colour Interaction in LEPTO decreases the cross 
section in the smallest \xbj\ bin by about 50\%, 
but does not affect the large--\xbj\ region.
Increasing the probability of Soft Colour Interaction from the default value
of 50\% to 100\% does not increase the cross section.

The region of large \xbj\ can be seen more clearly in figure~\ref{had_ari}b).
The data, ARIADNE and LEPTO converge at larger \xbj.
In this region, where $x_{Jet}$ approaches \xbj, the phase
space for parton emission is small. 
Therefore, the cross section is expected to be largely
independent of the parton shower mechanism. 
On the other hand, HERWIG and LDC stay below the data. 

The excess of forward jets at small \xbj\ observed in the data with 
respect to LEPTO and HERWIG may be interpreted as an indication of hard 
physics not implemented in present models of DGLAP--based parton evolution.
However, the current implementation of BFKL--type physics, 
as exemplified by the LDC model, 
still underestimates the measured forward jet cross section.
\vspace*{1.cm}

\begin{table}[htb]
\renewcommand{\arraystretch}{1.8} 
\begin{center}
\begin{tabular}{|r@{ -- }l||c|c|}
\hline
\multicolumn{2}{|c||}{$x$  range} &
$\frac{\displaystyle d\sigma}{\displaystyle dx}
\pm{\rm stat.}\pm{\rm syst.}~[{\rm nb}]$ & 
syst. ($E_{CAL}$ scale) $[$nb$]$\\
\hline \hline
$4.5\cdot10^{-4}$ & $8.0\cdot10^{-4}$ & $114 \pm 9.7 ^{\,+\,29}_{\,-\,15}$ 
& ($-5.9,~+18$) \\
$8.0\cdot10^{-4}$ & $1.4\cdot10^{-3}$ & $96.2 \pm 6.5 ^{\,+\,8.2}_{\,-\,8.2}$ 
& ($-8.1,~+7.8$) \\
$1.4\cdot10^{-3}$ & $2.5\cdot10^{-3}$ & $77.8 \pm 4.7 ^{\,+\,5.2}_{\,-\,6.9}$ 
& ($-4.2,~+7.0$) \\
$2.5\cdot10^{-3}$ & $4.5\cdot10^{-3}$ & $34.4 \pm 2.2 ^{\,+\,3.8}_{\,-\,1.9}$
& ($-2.1,~+2.6$) \\
$4.5\cdot10^{-3}$ & $8.0\cdot10^{-3}$ & $14.1 \pm 1.0 ^{\,+\,2.5}_{\,-\,1.2}$
& ($-1.2,~+1.3$) \\
$8.0\cdot10^{-3}$ & $1.4\cdot10^{-2}$ & $6.53 \pm 0.54 ^{\,+\,0.1}_{\,-\,0.7}$
& ($-0.7,~+0.2$) \\
$1.4\cdot10^{-2}$ & $2.5\cdot10^{-2}$ & $2.65 \pm 0.25 ^{\,+\,0.3}_{\,-\,0.3}$
& ($-0.03,~+0.2$) \\
$2.5\cdot10^{-2}$ & $4.5\cdot10^{-2}$ & $0.65 \pm 0.09 ^{\,+\,0.1}_{\,-\,0.4}$
& ($-0.00,~+0.05$) \\
\hline
\end{tabular}
\vspace*{0.5cm}
\caption{\it Forward jet cross sections and their errors for the 
kinematic region given in table~\ref{cuts}. The last column shows the 
correlated systematic error due to 
the energy scale uncertainty of the calorimeter, which is not included in the
central column. It corresponds to the shaded band in figure~\ref{had_ari}.}
\label{tab_results}
\end{center}
\end{table}

\clearpage

\section{Summary and Conclusions}
\label{summary}
An investigation of forward jet production including a comparison 
to various Monte Carlo models has been performed.
Three regions are identified in the $E^2_{T,Jet}/Q^2$ distribution:
i) the standard DGLAP region with $E^2_{T,Jet} \ll Q^2$, where all 
Monte Carlo models are in agreement with the data;
ii) the region of phase space where BFKL dynamics is expected 
to contribute 
significantly with $E^2_{T,Jet} \approx Q^2$, where only the Colour 
Dipole model describes the data well, and
iii) the region with $E^2_{T,Jet} \gg Q^2$,
where none of the models reproduces the data.

The forward jet cross section at hadron level is measured in the region 
ii) where $E^2_{T,Jet} \approx Q^2$.
The cross section is compared to the predictions of several models:
ARIADNE, which includes one of the main features
of the BFKL--based phenomenology, that is the absence of the strong
ordering in the transverse momenta in the parton shower; 
LDC, which is based on the CCFM approach 
and which smoothly interpolates between the BFKL and the DGLAP predictions 
in their range of validity; and, 
LEPTO and HERWIG, which are based on leading order DGLAP parton evolution.
The measured cross section is reasonably well described by ARIADNE
while LEPTO, HERWIG and LDC predict cross sections that
are too low at small $x$. 
The excess of forward jets at small \xbj\ observed in the data with respect 
to LEPTO and HERWIG may be interpreted as an indication of hard physics 
not implemented in present models of DGLAP--based parton evolution.
However, the current implementation of BFKL--type physics, as exemplified 
by the LDC model, still underestimates the measured forward jet cross section.

\section*{Acknowledgements}
We thank the DESY directorate for their strong support and encouragement. 
The remarkable achievements of the HERA machine group were essential for the
successful completion of this work and are gratefully acknowledged. We also 
thank J.~Bartels, G.~Ingelman, L.~L\"onnblad, E.~Mirkes and M.~W\"usthoff 
for many useful discussions and M.~W\"usthoff for providing the BFKL 
theory curves.




\end{document}